\documentclass[review,pdftex]{elsarticle}
\usepackage{lineno}
\usepackage{color}
\usepackage{longtable}

\usepackage{subfigure}

\usepackage[inline]{enumitem}

\usepackage{bm}
\usepackage{amsmath}
\usepackage{amssymb}
\usepackage{txfonts}
\usepackage{url}

\usepackage{cleveref}

\modulolinenumbers[5]
\usepackage{booktabs}

\journal{Surface and Interfaces}

\begin{document}
\begin{frontmatter}
  \title{
      {\it Ab-initio}-based Interface Modeling and Statistical Analysis
      for Estimate of the Water Contact Angle on a Metallic Cu(111) Surface
  }

\author[1]{Takahiro Murono\corref{cor1}}
\ead{tak.murono@icloud.com}

\author[2]{Kenta Hongo\corref{cor1}}
\ead{kenta\_hongo@mac.com}

\author[1]{Kousuke Nakano}

\author[1]{Ryo Maezono}

\cortext[cor1]{Correspoding authors}

\address[1]{School of Information Science, JAIST}
\address[2]{Research Center for Advanced Computing Infrastructure, JAIST}

\begin{abstract}
  Controlling the water contact angle on a surface is important
  for regulating its wettability in industrial applications,
  which involves developing {\it ab initio} prediction scheme of accurately predicting the angle.
  The scheme requires structural models for the adsorption of liquid molecules on a surface,
  but their reliability depend on whether the surfaces comprise insulating or metallic materials.
  Previous {\it ab initio} studies have focused on the estimation of the water contact angle on insulators,
  where the periodic-honeycomb array of water molecules
  was adopted as the adsorption model for the water on the insulating surface
  and succeeded in the insulating cases.
  This study, however, focus on the water contact angle on a metallic surface,
  and propose a simple {\it ab initio} based estimation scheme.
  We not only adopt the previously proposed structural modeling
  based on the periodic-honyecomb array, but also consider
  an ensemble of isolated water oligomers that have different
  molecular coverage (ML) values.
  We established a statistic model to predict a contact angle of
  the water wetting on a Cu(111) surface:
  The coverage-dependent contact angles
  obtained from each of the isolated clusters
  was fit to a quadratic regression,
  and the contact angle was interpolated
  by referring to a ML value of water layer in literature.
  This interpolated value lay within the deviation of experimental angles.
  In addition, the Boltzmann-average over the isolated clusters
  was found to agree well with the interpolated one.
  This indicates that the Boltzmann-average is useful
  for estimating the contact angle of other metallic surfaces
  without knowing a ML value {\it a priori}. 
\end{abstract}
\end{frontmatter}

\section{Introduction}
Regulating the wettability of surfaces~\cite{2009LAN,2016CAR,2017HON,2019TAK}
is an important issue that must be addressed to broaden their industrial applications,
such as heterogeneous catalysis, corrosion, 
and electrochemistry.~\cite{2016CAR,1987THI,2002HEN} 
The wettability of the surfaces of metal catalysts such as Cu has a significant effect on 
their catalytic activity 
for electrochemical reactions,~\cite{2015STO}
wherein the receding contact angle of water on the metal surface corresponds to the 
oxygen reduction reactions of the catalyst. 
This indicates that the efficiencies 
of the electrochemical reactions of metal catalysts
can be enhanced by controlling the water contact angle with the metal surface. 

\par
The contact angle ($\theta$)
is the primary measure 
of wettability; a larger contact angle ($\theta > 90^\circ$) indicates hydrophobicity,
i.e., low wettability,
whereas a smaller contact angle ($0^\circ < \theta < 90^\circ$) indicates hydrophilicity,
i.e., high wettability. 
Experimentally, the water contact angle on metal surfaces can be measured 
directly by observing the angle 
captured by the cameras.~\cite{2013YUA,2018HUH} 
The angle is determined using the Young's relation~\cite{1805YOU,1985GEN}:
\begin{equation}
\cos\theta = \frac{\gamma_{\rm{SG}} 
- \gamma_{\rm{SL}}}{\gamma_{\rm{LG}}}
\ ,
\label{eq:Young-2}
\end{equation}
where 
$\gamma_{\rm{SG}}$, $\gamma_{\rm{LG}}$, 
and $\gamma_{\rm{SL}}$ are 
the surface energies at 
solid--gas, liquid--gas, 
and solid--liquid interfaces, respectively.
Although these energies can be computed by various approaches,
the proposed study focuses on {\it ab initio} evaluations
(the comparison with other approaches is provided later).
The {\it ab initio} prediction of water contact angles on surfaces
was recently introduced, and it has been successfully
applied to estimate this angle 
on the surfaces of insulators.~\cite{2009LAN,2016CAR} 
For example, the contact angle of water on Si~\cite{2009LAN} 
was predicted via {\it ab initio} evaluation ($88^\circ$),
which was close to the experimental value ($91^\circ$).
In addition, {\it ab initio} evaluations have also been applied to estimate
the water contact angle on transition metal oxides~\cite{2016CAR}. 
For example, the water contact angle on CeO$_2$(111) [Nd$_2$O$_3$(0001)]
was estimated via {\it ab initio} evaluations as $100^\circ$ [$103^\circ$],
which is comparable to the experimental value, i.e., $103^\circ \pm 2^\circ$ [$101^\circ \pm 3^\circ$];
this result indicates that the {\it ab initio} evaluations are in accordance with actual values. 
For the estimations performed in previously reported studies, 
the $\gamma_{\rm{LG}}$ of water was substituted with that of crystal ice
because water and crystal ice have similar surface energies~\cite{1992VAN}. 
More sophisticated {\it ab initio} models based on quantum molecular dynamics (QMD) 
have been proposed and applied to several insulating systems
such as CaCO$_3$~\cite{2017JYL_TJZa,2019JYL_TJZa}
and graphite~\cite{2018JYL_MCa};
for CaCO$_3$ and graphite, 
their comptational values were $38^\circ$ and $70^\circ$, respectively,
which agree well with experimental values of $41^\circ \pm 4^\circ$
and $68^\circ\pm 2^\circ$, respectively.

\par
\begin{figure}
  \begin{center}
    \includegraphics[width=8cm]{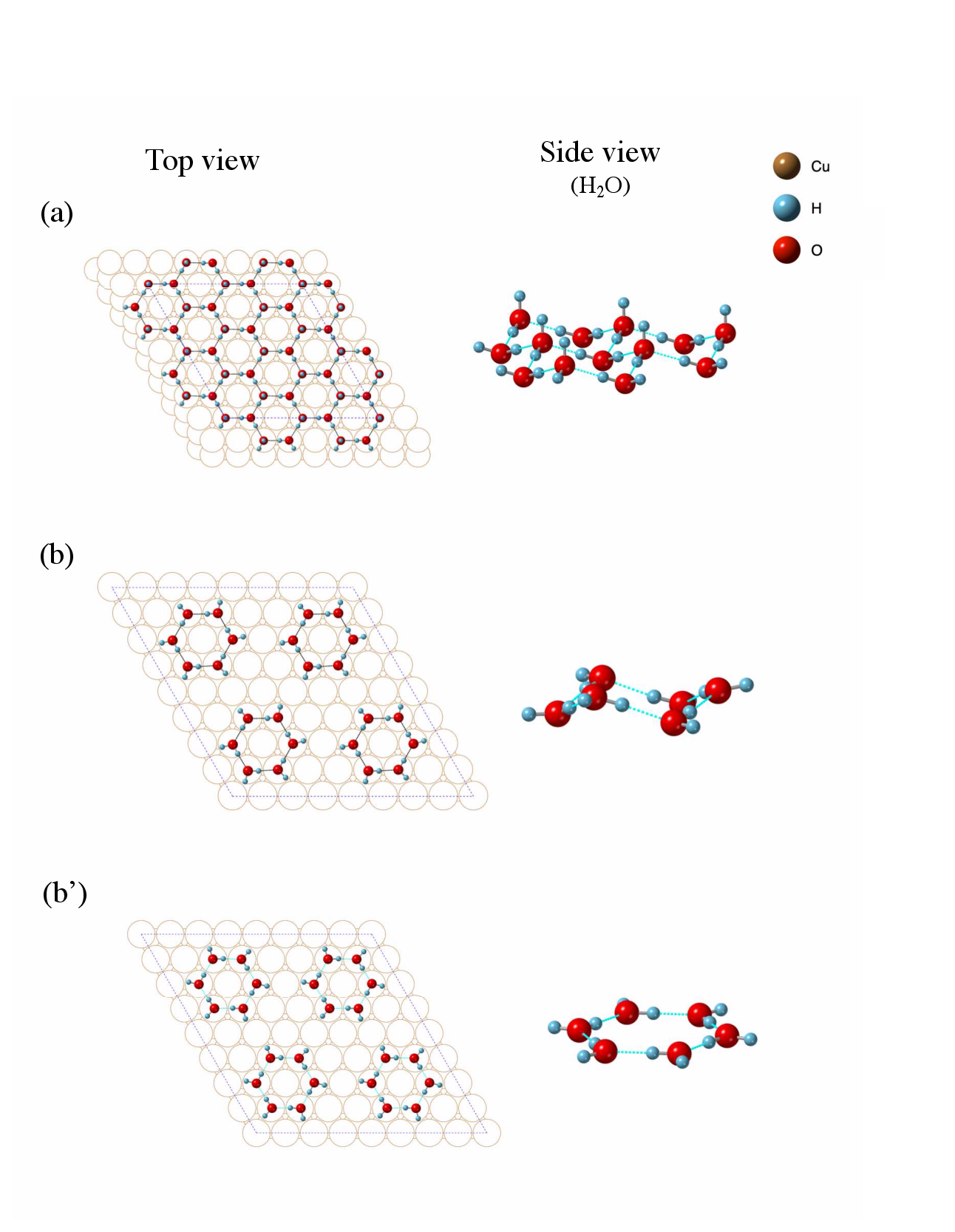}
    \caption{
      Structural models describing the adsorption of water on metal surface for 
the {\it ab initio} evaluation of 
the contact angles. 
Periodic honeycomb structure 
known as the ``bilayer model.''~\cite{2009HOD,2012CAR}
Panel (a) shows the bilayer model, which is used for the 
estimation of the contact angles of water on insulator surfaces.~\cite{2009LAN,2016CAR} 
Panel (b) and (b') show 
the ``isolated molecular (cluster) models,''
which were found to be more appropriate for 
the estimation of the water contact angle on metal surfaces. 
    }
    \label{fig.strComp}
  \end{center}  
\end{figure}

Previously reported studies based on the {\it ab initio}
evaluations of contact angles of water at different surfaces
have considered insulators as the surfaces. 
In previous studies,~\cite{2016CAR,2009LAN}
an ``ice-like bilayer model''~\cite{2009HOD,2012CAR}
(periodic-honeycomb structure of ice)
has been adopted as the adsorption model for
conducting the {\it ab initio} evaluations of the contact angles of
water on various insulator surfaces,
which is a natural choice for the $\gamma_{\rm LG}=\gamma_{\rm ice}$
model (see Fig.\ref{fig.strComp} (a)).
Note that the experimental contact angles for the insulating surfaces
are definitive due to their small errors and hence
the ice-like bilayer modeling can be validated by
comparing the resultant values with the experiments.
Similarly, the QMD-based approaches can be validated,
even though their computation is heavily demanded.

\par
However, when estimating the water contact angle on the surface of metals, 
the structural modeling to describe 
the adsorption of water molecules on the metallic surface 
should be different from that used
to describe the adsorption of water molecules on 
the insulator surface. 
Previous studies have reported the presence of isolated water hexamers
on the surface of metals such as Cu~\cite{2002MOR,2007MICb},
implying that the ``bilayer model'' would not be necessarily appropriate for
the {\it ab initio} evaluation of the water contact angle on the metal surface. 
Instead, ``isolated models'' that do not exhibit the bridging of hydrogen bonds
over neighboring unit cells would be rather appropriate. 
In addition, experimental values of
the water contact angle on the Cu surface 
deviate from $60^\circ$ to $85^\circ$,
depending on various experimental conditions
that are not completely controllable.~\cite{2015EO_GKa,2015PAR,2008VOI}
This implies that unlike the insulating surface,
the validation of theoretical works
on the water-copper wettability is difficult 
as mentioned before,
but only the range of the contact angles
can be addressed.

\par
In this study, we aim at establishing the {\it ab-initio}-based
scheme of estimating the water contact angle on a metallic copper surface,
i.e., the metallic Cu(111) surface.
We first construct several isolated models (cluster/oligomer) and then
peform the {\it ab initio} adsorption energy evaluation of
the cluster models as well as that of the bilayer model.
We observed that 
the type of $N$-mer/oligomer (ranging from monomer to hexamer) and 
number of $N$-mers within a unit cell should be specified. 
Particularly, we examined the coverage-dependent contact angles,
which is equivalent to the dependence of the adsorption energy
on the size/coverage of the oligomers.
Referring to a reported coverage, an interpolation between
the oligomer and periodic models leads to our {\it ab initio}
prediction of the water-copper contact angle.
We found that (i) the coverage-dependent contact angles
monotonously decreases as the coverage increases,
(ii) deviation of the coverage dependence is less than
that of the experimental angles ($25^\circ$),
and hence (iii) the predicted contact angle 
lies between the experimental deviation.
For comparison, we considered the Boltzmann average
over the individual contact angles,
which was found to be similar to the contact angle
predicted from the coverage-dependent angles.

\par
The paper is organized as follows.
The ``Computational Methods'' is divided into two subsections.
The ``Interface modeling'' subsection provides a detailed description of
our ``isolated cluster'' models.
The ``Computational details'' subsection
specifies our {\it ab initio} methods of evaluating the energies
based on the models.
The ``Results and Discussion'' section shows
the numerical results obtained from our approach,
and deals with {\it pros} and {\it cons} of our approach
by comparing the proposed approach with other approaches.
Finally, the findings are summarized in the ``Conclusion'' section.

\section{Computational Methods}
\subsection{Interface modeling}
Based on the previous studies,~\cite{2009LAN,2016CAR}
we estimated
\begin{equation}
  {\gamma _{{\rm{SG}}}}-{\gamma _{{\rm{SL}}}}
  =  -{{E_{{\rm{ads}}}^{\rm{water}}}}/{A}
\ ,
\label{eq:tm.05.27.18.03}
\end{equation}
where $E_{\rm{ads}}^{\rm{water}}$ 
is the adsorption energy 
of water molecules on the surface 
($A$ is the unit area within which 
$E_{\rm{ads}}^{\rm{water}}$ is defined). 
For the liquid-gas interfaces, 
\begin{equation}
  \gamma_{\rm LG}=\gamma _{{\rm{ice}}}
\ ,
\label{eq:tm.12.26.19.01}
\end{equation}
which was evaluated using the surface 
energy of ice, $\gamma_{\rm ice}$.~\cite{2009LAN,2016CAR} 
Then, the formula for the contact angle can be provided as follows: 
\begin{equation}
  \cos \theta
  = \frac{{ - E_{{\rm{ads}}}^{{\rm{water}}}}}
  {{A\cdot {\gamma _{{\rm{ice}}}}}}
\ , 
\label{cosAbinit}
\end{equation}
which can be reduced to the {\it ab initio} 
energy evaluations. 
$E_{\rm{ads}}^{\rm{water}}$ is given as~\cite{2016CAR}:
\begin{equation}
  E_{{\rm{ads}}}^{{\rm{water}}}
  = {E_{{\rm{tot}}}} - {E_{{\rm{water}}}}
  ({\rm{onSurf}}) - {E_{{\rm{CuSlab}}}}
  \ , 
\label{iceAbs}
\end{equation}
where $E_{\rm{CuSlab}}$, 
$E_{\rm{water}}({\rm{onSurf}})$, 
and $E_{\rm{tot}}$, 
are the energy of 
the Cu slab, energy of 
water molecules on the surface of 
the adsorbed structure, and total energy 
of the system with the molecules 
on the surface, respectively. 
The surface energy $\gamma_{{\rm{ice}}}$ 
can be evaluated as: 
\begin{equation}
  \gamma _{{\rm{ice}}}
  =\cfrac{E_{\rm{ice}}(n;{\rm Slab})
  -E_{\rm ice}(n;{\rm Bulk})}{2A_{\rm{ice}}}
\ , 
\label{gammaIce}
\end{equation}
which is a measure of the stabilization 
of the ice surface, 
where $E_{\rm{ice}}(n;{\rm Slab})$ 
and $E_{\rm ice}(n;{\rm Bulk})$ 
are the energies of the slab and 
the bulk of the ice composed of 
$n$ water molecules, respectively. 

\par
The water contact angle on the surface
is estimated using five quantities: 
$E_{\rm{ice}}(n;{\rm Bulk})$, 
$E_{\rm{ice}}(n;{\rm Slab})$, 
$E_{\rm{CuSlab}}$, 
$E_{\rm{water}}({\rm{onSurf}})$, 
and 
$E_{\rm{tot}}$, which can be evaluated 
using {\it ab initio} density functional theory (DFT) 
calculations. 
Considering Cu(111) as the surface, 
we calculated the energies 
using a DFT package, CASTEP.~\cite{2005CLA}
For the ease of comparison with previous works,~\cite{2009LAN,2016CAR} 
we used the same exchange-correlation 
function, GGA-PBE,~\cite{1996PER} 
used in previous studies.
Because the valence shell of Cu consists of $d$-electrons, 
the Hubbard $U$ correction~\cite{2020HAN,2019ICHc,2018GHA} 
may also be important. 
However, this correction matters  
when the orbital becomes more localized 
where the self-interaction 
cancellation is damaged.~\cite{1991ANI}
For our metallic Cu surface, 
the orbital is delocalized, and hence, 
we did not consider such corrections, 
as in most previous studies.~\cite{2014MAK}
In addition, we did not consider the exchange-correlation
functional with van der Waals (vdW) corrections
due to Tkatchenko and Scheffler~\cite{2009TKA},
because the vdW-DFT is know to fail to reproduce 
the delocalized nature of electrons.~\cite{2017JYL_TJZa,2013TB_JGAa}
Accurate reproduction of the non-covalent
molecular interactions 
involves other approaches such as
quantum Monte Carlo (QMC),~\cite{2013HON,2019CRH_CMW}
but QMC is not applicable to
our target system including water clusters on
the delocalized metallic surface
due to its much heavier computation than DFT.
Norm-conserving pseudopotentials~\cite{1979HAM} 
were used to describe the ionic cores. 
The detailed computational conditions 
for each calculation is summarized 
in Table~\ref{tab:DFT_condition}.

\par
The detailed information used for modeling the geometries of 
the adsorbed molecules is provided in the 
subsequent subsection: Computational Details. 
The main comparison was conducted between the predictions achieved by 
the periodic-honeycomb model (bilayer)~\cite{2009HOD,2012CAR} 
[panel (a) in Fig.~\ref{fig.strComp}] 
and those by the isolated molecular models 
(buckled)~\cite{2007MICb,2007MICa} [panel (b)], 
which were evaluated using a Cu slab 
comprising nine atomic layers. 
Furthermore, to investigate the existence of the considerable bias on the choice 
of $N$-mer and coverage, we compared 
the predictions with $N=1,2,3,4$, and $6$;
however, this comparison was conducted with reduced cost and complexity of the computation, 
namely with an H-parallel model~\cite{2007TAN}
(planar model) [panel (b') in Fig.~\ref{fig.strComp}]
and with reduced number of layers (four). 
$N = 5$ was excluded because we limited the 
possible geometry to satisfy 
the ``on-top alignment,''
which is supported by the findings of previous studies~\cite{2007MICa,2007TAN} 
[i.e., $N=5$ cannot accommodate 
the molecules within a unit cell, such that 
they have an ``on-top alignment'' geometry]. 
For the oligomers, $N>2$, 
they can either assume the chain or circular form. 
Therefore, we considered both possibilities for $N = 3$ and $4$
and considered only the circular form for $N = 6$, 
as supported by previous studies.~\cite{2007MICb,2007MICa}
The dependence on the coverage,
i.e., on the number of $N$-mers 
placed within a unit cell, was 
examined up to $N = 3$. 

\subsection*{Computational details}
For the energies required to evaluate 
the contact angles, 
we prepared 
the geometries of 
the Cu slab~(for $E_{\rm{CuSlab}}$), 
ice bulk slab~(for $E_{\rm{ice}}(n;{\rm Bulk})$ and $E_{\rm{ice}}(n;{\rm Slab})$), 
and the water molecules 
adsorbed on the Cu slab
~(for $E_{\rm{water}}({\rm{onSurf}})$ and $E_{\rm{tot}}$). 
\begin{table}
 \begin{center}
   \caption{
   Computational details of the DFT 
   calculations for each energy 
   to evaluate the contact angles of water on
   the metallic Cu(111) surface. 
   The $k$-mesh and $E_{\rm{CUT}}$ 
   (given in [$\rm{eV}$]) are the 
   mesh size for the Brillouin zone~\cite{1976MON} 
   and the plane wave energy cutoff, 
   respectively. 
   These values were determined 
   by the convergence of the total energies. 
   'Cu-struct-opt' and 'Cu-surf-relax' 
   indicate the optimized Cu bulk and Cu-slab structures, 
   respectively. 
     } 
     \label{tab:DFT_condition}
     \begin{tabular}{lcc} 
       \hline
           & $k$-mesh & $E_{\rm{CUT}}$ [eV] 
       \\
       \hline
       $E_{\rm{ice}}(n;{\rm Bulk})$   &  4$\times$2$\times$2   &  750 \\
       $E_{\rm{ice}}(n;{\rm Slab})$   &  4$\times$4$\times$1   &  750 \\
       $E_{\rm{CuSlab}}$              &  1$\times$1$\times$1   &  700 \\
       $E_{\rm{water}}({\rm{onSurf}})$  &  1$\times$1$\times$1   &  600 \\
       $E_{\rm{tot}}$                 &  1$\times$1$\times$1   &  700 \\
       \hline
       Cu-struct-opt                &  8$\times$8$\times$8   &  800 \\
       Cu-surf-relax                &  11$\times$11$\times$1 &  800 \\
       \hline
     \end{tabular}
 \end{center}
\end{table}

\par
A Cu bulk with an initial lattice constant of 
$3.6147$~\AA{} was prepared to 
construct the Cu-slab structure. 
The constant was optimized under the 
bulk structure as $3.728$~\AA{}; 
subsequently, 
the bulk was cleaved along the (111) plane 
and used as the initial structure of the slab. 
Then, a nine(four)-layered slab was located 
periodically, on which a $30$~\AA{} deep 
vacuum layer was attached; the atomic positions within 
the three (one) layers from the surface 
were relaxed by optimizing the geometry
to obtain the final structure of the Cu slab. 
For the ice bulk, 
we used the Ih structure 
with a lattice constant of 
$4.516$~\AA{}.~\cite{1958LON,1961CHI} 
The ice bulk was cleaved to obtain 
four layers of H$_2$O, 
which were parallel to 
the basal plane of the Ih crystal. 
The Cu slab was formed by attaching 
a vacuum layer with a 
$30$~\AA{} depth, without 
further relaxations. 

\par
The water molecule was prepared 
using the initial geometry, 
$l_{\rm{O-H}} = 0.96~$~\AA{} and 
$\theta_{\rm{H-O-H}} = 104.5^\circ$.
For the periodic-honeycomb model
[panel (a) in Fig.~\ref{fig.strComp}], 
we used the ``bilayer model'' employed 
in preceding studies.~\cite{2009HOD,2012CAR}
The structure was generated 
from the ``H-parallel'' structure,~\cite{2007TAN} 
where all the H$_2$O molecular planes 
were parallel to the slab surface. 
Then, the positions of the oxygen atoms were 
adjusted such that 
their heights from the slab surface 
were alternated in every molecule 
to form the buckled structure.~\cite{2007MICb,2007MICa}
To evaluate the contact angles in the four-layer slab models
(as shown later in Fig.~\ref{fig2}), 
water molecules were placed on the models using
the on-top alignment geometry~\cite{2007MICa,2007TAN} 
with the H-parallel orientation~\cite{2007TAN}; 
herein, the vertical height was optimized by DFT.

\section{Results and Discussion}\label{results_discussion}
\subsection{Coverage-dependent contact angles}
\begin{figure}[h]
\begin{center}
\includegraphics[width=120mm,clip]{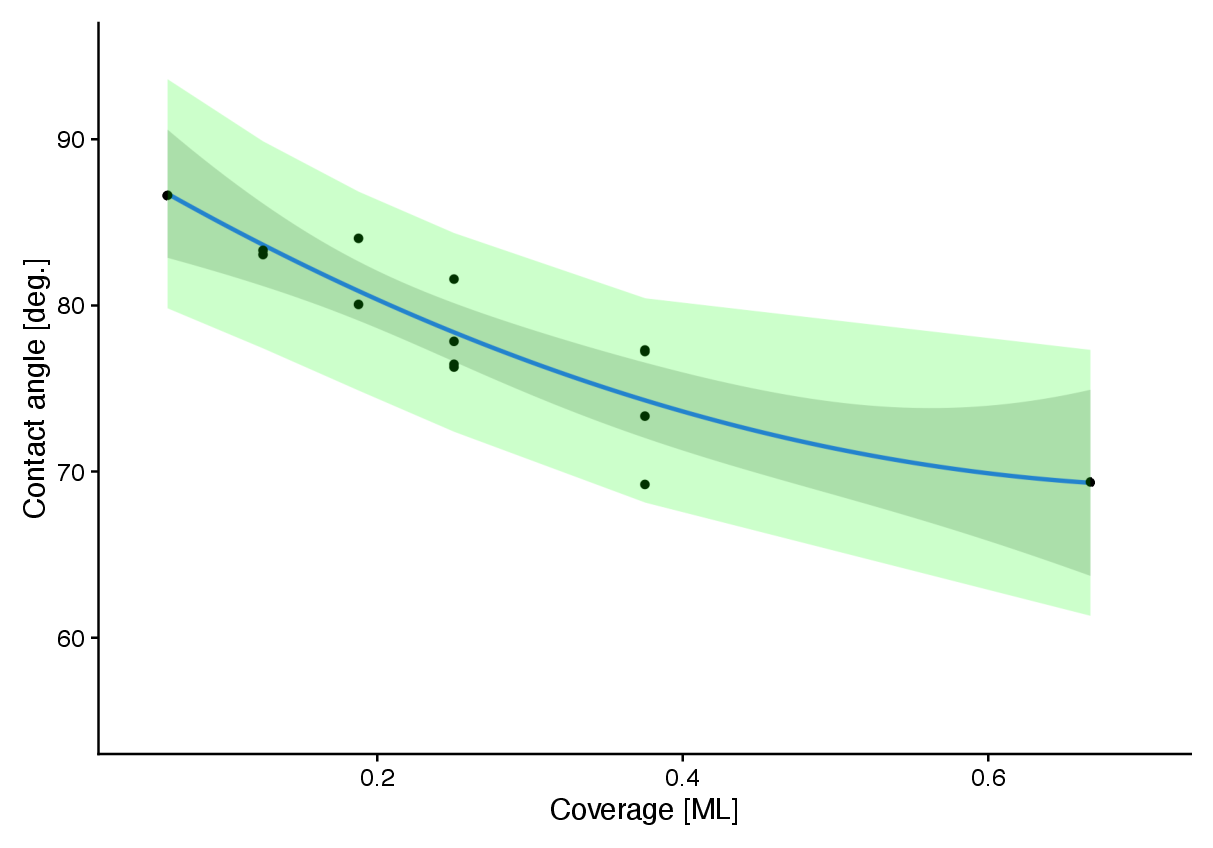}
\end{center}
\caption{
Dependence of the predicted contact angles 
on coverage values based on
various adsorption models for the adsorption
of water molecules on 
the metallic Cu(111) surface.
Blue line indicates a quadratic regression;
gray and green areas correspond to 
95\% confidence and prediction intervals, respectively;
the (adjusted) $R^2$ value is $0.73$.
}
\label{fig2}
\end{figure}
Figure~\ref{fig2} shows the coverage-dependent contact angles,
together with a quadratic regression that is the best fit
in terms of both $R^2$ and AIC within quartic polynomial (see Supporting Information).
The largest coverage of $6/9\sim 0.667$ ML corresponds to the bilayer model,
whereas the smallest one of $1/16=0.0625$ ML
corresponds to the isolated monomer model;
each of them is a unique pattern and hence has no error bar.
In contrast, in their intermediate coverage range, 
some structural patterns of different oligomers exhibit
an identical coverage value, thereby deviating;
the possible patterns of isolated models increase
as the coverage increases: 
[monomer, dimer] at $2/16=0.125$ ML,
[monomer, trimer(chain)] at $3/16=0.1875$ ML,
[monomer, dimer, tetramer(chain)] at $4/16=0.25$ ML, and 
[dimer, trimer(ring/chain), hexamer(ring)] at $6/16 = 0.375$ ML.

\par
It is found from Fig.~\ref{fig2} that
the coverage-dependent contact angles
monotonously decreases as the coverage increases.
In order to determine the computed contact angle from the dependence,
we need to determine an appropriate value of coverage
for water layer on Cu(111).
The knowledge of the coverage enables us to estimate the contact
angle from the coverage-dependence of the angle.
In the present study, we adopt the coverage of $\sim 0.4$ ML
that is estimated from a previous study~\cite{2016ZHU}
based on a Langmuir adsorption isotherm.~\cite{1918LAN}
Referring to the fitted curve, an interpolation into the $0.4$~ML 
leads to the contact angle of $74 \pm 6^\circ$.
As can be seen from Table~\ref{tab.result01},
the predicted/interpolated contact angle at $\sim 0.4$ ML ($74 \pm 6^\circ$)
is same as the contact angle of isolated cluster model at $6/16(= 0.375)$~ML
($74 \pm 6^\circ$) within the present significant figures.
This implies that even the isolated cluster models having
the coverage closest to the water layer 
makes a good approximation.
In addition, its 95\% prediction interval ($[67,80]$) lies
within the range of the experimental angles ($[60,86]$).
\begin{table*}
  \begin{center}
    \caption{
     Predicted water contact angles 
    on a Cu(111) surface 
    using different absorption 
    models.
    ``Boltzmann ave.'' implies 
    the average of several 
    isolated models with different 
    sizes of isolated molecules, 
    providing varying coverage 
    (see text).
    }
    \label{tab.result01}
    \begin{tabular}{cc}
      \hline
      Model &  Contact angle [$^\circ$] \\
      \hline
      Periodic-honeycomb (bilayer) &  69 \\
      Isolated model at $0.375$~ML & $74\pm 6$ \\
      Predicted at $0.4$~ML &  $74\pm 6$ \\ 
      Boltzmann average at 297 K& $78 \pm 5$ \\      
      \hline
       Exp.~\cite{2015PAR} & 86 \\
       Exp.~\cite{2008VOI} & 75 \\
       Exp.~\cite{2015EO_GKa} I & $77\pm 7$ \\
       Exp.~\cite{2015EO_GKa} II & $70\pm 8$ \\
       Exp.~\cite{2015EO_GKa} III & $70\pm 10$ \\              
       \hline
    \end{tabular}
  \end{center}
\end{table*}
\par
Compared to the isolated-cluster-based contact angles
and the interpolated angle,
the periodic-honeycomb model (bilayer)~\cite{2009LAN,2016CAR}
gives a smaller contact angle ($69^\circ$),
though it lies within the range of experimental angles.
The difference between the honeycomb model and
the others cannot be 
attributed to the lattice mismatch
between the Cu surface and honeycomb network,
because the lattice mismatch between 
the Cu surface and honeycomb network was $-2.1\%$~\cite{2009HOD}.
The coverage-dependence of contact angles 
can be attributed to the bridging of the hydrogen 
bonds over neighboring unit cells.
The periodic-honeycomb model can be regarded
as differing from a realistic interface assembling
water clusters in terms of the coverage-dependence.

\par
In addition, we evaluate the Boltzmann-averaged contact angle
over the isolated cluster models,
leading to its contact angle of $78\pm 5^\circ$.
The Boltzmann-averaged angle agrees well with
both the isolated-cluster-based and interpolated ones,
though the former is slightly larger than the latter.
The Boltzmann-average scheme
is a straightforward way of evaluating the contact angle
by considering various isolated cluster models,
but taking a less computational cost than investigating
the coverage estimation.~\cite{2016ZHU}
Thus, the Boltzmann-average scheme is applicable to other metallic surfaces
even if their appropriate coverage values are {\it a prior} unknown.

\par
Further, we examined the 
reliability of results by comparing the intermediate values
obtained using Eq.~\eqref{cosAbinit},
$\gamma_{\rm{ice}}$, and $E_{\rm{ads}}^{\rm{water}}$,
with those obtained in the previous studies. 
The surface energy value obtained in this study 
($\gamma_{\rm{ice}} = 41.83$ [meV/\AA$^2$])
is comparable to that reported in a previous DFT study 
($44$ [meV/\AA$^2$]),~\cite{2016CAR} 
indicating a fair coincidence. 
In addition, the adsorption energy estimated in this study 
($E_{\rm{ads}}^{\rm{watar}} = 238 $ [meV/H$_2$O] (monomer)) is lower than
that of the experimental value of Cu(111)
($352$ [meV/H$_2$O] $\approx 34$ [kJ/mol])
obtained using a temperature-programmed desorption evaluation
(TPD).~\cite{2007TAN} 
The apparent underestimation of our value can be attributed to 
the fact that the experimental values consider the energy required 
to break the hydrogen bonds during the evaporation of water molecules,
thereby resulting in the more stabilized value. 
The same explanation is applicable to
the DFT values (for monomer) obtained in our study,
which are smaller than those obtained in previous studies
($145-157$ [meV/H$_2$O]~\cite{2007MICa} and 
$187$ [meV/H$_2$O] $\approx 18$ [kJ/mol])~\cite{2007TAN}). 
In addition, the DFT values for dimers in previous studies 
($321-332$ [meV/H$_2$O]~\cite{2007MICa} and 
$352$ [meV/H$_2$O] $\approx 34$ [kJ/mol])~\cite{2007TAN})
were larger than those obtained in this study.
The significant increase in values for the dimer
can be attributed to the difference in definition. 
In their definition, 
the hydrogen bonding interactions 
between water molecules were 
included in the $E_{\rm{ads}}^{\rm{watar}}$, 
and hence, it increased 
with an increase in the size of $N$-mer. 

\subsection{Water coverage on metallic surfaces}
Here we make some remarks on the water coverage that is a key quantity
of estimating the contact value from the coverage-dependence,
though it is not necessary for the Boltzmann-average-based estimation.
As a matter of fact,
the Cu(111) surface has a lower coverage than other metallic surfaces,
{\it e.g.}, Pd(111), Pt(111), and Ru(0001).~\cite{2014REV,2016BEL} 
This indicates that the isolated model is appropriate for the former,
whereas a network of interacting water molecules is important for the latter.
The lower coverage of Cu(111) is attributed
to the weaker adsorption of molecules on Cu(111),
namely, $\sim$ 352 [meV/H$_2$O],
as estimated by TPD experiments,~\cite{2007TAN} 
which is approximately half of adsorption energy
on Pd, Pt, and Ru.~\cite{2014REV} 
The molecules adsorbed on Cu then more easily desorb again from the surface,
leading to the lower coverage than those for Pd, Pt, and Ru.

\par
Corresponding to the lower coverage, 
isolated monomers, dimers, trimers, and 
tetramers have actually been observed
in STM (scanning tunneling microscope) experiments 
on several metallic surfaces.~\cite{2012CAR}
The lower coverage for Cu(111) was derived in a previous study~\cite{2016ZHU} 
by using a Langmuir adsorption isotherm.~\cite{1918LAN}
The dependence on the adsorption energy 
appears through the adsorption equilibrium constant 
as a function of the Boltzmann factor. 
Because the Langmuir adsorption isotherm~\cite{1918LAN} 
is basically applicable to 
solid--gas (steam) interfaces, 
applying it to a solid--liquid interface requires careful consideration. 

\par
Applying it in this case would be justified when 
(i) the intermolecular interactions 
between water molecules are negligible, and 
(ii) the absorption process of the water molecules on the solid surface
is equilibrated and then none of the absorbed molecules
are dissociated into H$^+$ and OH$^-$,
so the molecules in the liquid water can then 
be regarded as molecules in steam. 
The isolated model satisfies (i), 
and the resultant lower coverage 
seems self-consistent. 
For (ii), the dissociation energy 
of the water molecules on Cu(111) 
has been reported to be 1.23~[eV/H$_2$O],~\cite{2010WAN} 
being much larger than the 
energy scale of the adsorption energy, 
$\sim$352~[meV/H$_2$O].~\cite{2007TAN}
A previous DFT study~\cite{2016ZHU} 
also concluded that the adsorption of water molecules 
on Cu(111) without dissociation 
is much more stable than that with dissociation. 

\par
In Eq.~\eqref{eq:tm.05.27.18.03} of the present study, 
the left-hand side denotes the balance of forces horizontal 
to the surface acting as the boundary of the liquid-covered region.
By achieving coverage, the system becomes stabilized by $E_{\rm{ads}}$, 
as indicated by the right-hand side of the equation. 
However, not only this stabilization term
but also another term, $\gamma_{\rm intMol}$, 
contributes to pulling the boundary inside toward liquid-covered region 
owing to intermolecular interactions. 
This tension term can be omitted when the intermolecular distance 
increases under the lower coverage, but when the coverage increases, 
the approximation in Eq.~\eqref{eq:tm.05.27.18.03} makes the 
the contact angle less accurate. 

\subsection{Comparison with other methodologies}
Herein, we compare the computational evaluations of the contact angle
obtained herein with those obtained previously
because it will be helpful for deeper understanding 
of the proposed scheme. 
The computational evaluations of the contact angle are
mainly classified into two schemes: 
(I) energetics combined with Young's relation in Eq.~\eqref{eq:Young-2} (indirect approach)
and (II) direct observations of droplet geometry (direct approach).~\cite{2018RAV,2019ESS,2019JIA}
Based on the scheme, several choices of interface modeling 
and computational methodologies are possible.
Hereafter, we discuss the universality of each approach.

\par
(I) The indirect approach first evaluates interface free energies by any means 
and then computes the contact angle via Eq.~\eqref{eq:Young-2}.
The free energies can be (approximately) evaluated from various simulations
including not only {\it ab initio} [electronic level] 
but also molecular dynamics (MD) and/or Monte Carlo (MC) [molecular level].
The accuracy of the contact angle depends on the adopted interface modeling 
and level of theory.

\par 
The {\it ab initio} simulations with interface modeling
appropriate for insulating surfaces (``ice-like bilayer'' model)
have proven to accurately reproduce the contact angles.~\cite{2009LAN,2016CAR}
Recently, a more sophisticated approach based on
{\it ab initio} molecular dynamics (AIMD) simulations
has been proposed by Lu {\it et al.} and applied
to insulators.~\cite{2017JYL_TJZa,2019JYL_TJZa,2018JYL_MCa}
The AIMD approach takes into account relaxation of
water molecules and then agree well with experiments.
This approach is, however, infeasible for metallic surfaces:
The AIMD simulation is done on the $\Gamma$-point,
so its simulation cell is larger than that of
usual band-structure simulations
in order to treat condensed matters.
Because of this requirement for a large supercell,
the AIMD approach is inadequate for our delocalized
Cu surface that requires a much larger supercell
than insulating surfaces.
Instead, as explained before,
the present study proposed ``isolated cluster'' models,
which relies on the experimental fact about a lower coverage
on the Cu surfaces.~\cite{2018SIM}
In addition, experimental contact angles on the Cu surface
deviate widely, depending on their sample preparations
and so on.
This indicates that we cannot make a simple comparison
between experiment and theoretical contact angles,
and hence validate reliability of our proposed interface modelling.
It is to be noted, however, that deviation of the coverage-dependent
contact angles is less than that of experimental angles.

\par
As for the molecular levels, several combinations of interface modeling and
the molecular simulations have been developed.
Within the framework of the molecular simulation, 
the interface energy is thermodynamically described
via fundamental quantities that can be evaluated by MD/MC sampling.
The phantom-wall method~\cite{2009LER,2010LER}, 
dry-surface method~\cite{2015LERb}, 
and interface wetting potential method~\cite{2015KAN,2017KAN} 
are combined with MD simulations;
the test-area perturbation approach~\cite{2005GLO,2007IBE} 
and the excess surface free energy approach~\cite{2008GRZ,2011RAN,2014KUM}
employ MC simulations.
As explained later, MD/MC requires molecular force fields
appropriate for individual metallic surfaces;
therefore, its applicability strongly depends on available
force fields.

\par
(II) The direct approach usually employs MD-based simulations to construct molecular droplets
and attempts to gauge the contact angles from the MD snapshots of the droplet geometries.
Several schemes~\cite{1999RUI,2003WER,2018RAV,2013SAN} 
have been proposed to identify the ``droplet shape" from the molecular condensation;
then, the contact angle is evaluated via elemental differential geometry.
Note that this direct approach (II) involves more uncertainty than the indirect one (I), 
but the direct one is applicable even for determining wettability
on rough surfaces and dynamic wettability.
This implies that (i) the indirect approach provides
more accurate (static) contact angle estimations
than the direct one, assuming the interface energies to be accurate~\cite{2019JIA},
and (ii) the direct approach provides more general estimations applicable to a wide range of 
wettability phenomena~\cite{2018RAV}.

\par
In both the direct and indirect approaches, 
the accurate prediction of the contact angle
strongly depend on that of molecular interactions at each interface.
Molecular force fields used in MD/MC simulations are crucial 
for reproducing the molecular interactions~\cite{2013TAH}, 
whereas {\it ab initio} DFT approaches usually predict the energies accurately.
In MD simulations, 
Lennard--Jones (LJ) potentials associated with electrostatic ones
are mostly selected as the force field.~\cite{2008ALE}
The LJ potentials require a set of parameters for each molecular interaction;
major parameter sets are available for water~\cite{2008ALE,2009VEG},
but individual parameter fittings are necessary for different surfaces by comparing with 
{\it ab initio} simulations or experiments.~\cite{1994ZHU, 2008HEI}
MD simulations with various types of the force fields have been applied
to evaluate water contact angles on metallic surfaces such as
Pt~\cite{2009SHI} and face-centered cubic Cu~\cite{2015XU},
but the accuracy of their predictions has been reported to
vary depending on the force fields adopted.~\cite{2003WER}

\par
Recently, several attempts have been made to address the force field issue in classical MD.
A straightforward way is to directly improve the force field accuracy
by fitting DFT data,~\cite{2013WU}
although its prediction should be calibrated by the experimental validation.~\cite{2003WER}
Further, {\it ab initio} MD (AIMD) has been applied to surface/interface properties of water and
evaluated the surface tension~\cite{2016NAG} and the contact angle~\cite{2012LI}.
The AIMD overcomes (semi)empirical force fields, but has a remarkable limitation on system sizes;
hundreds of water molecules in a previous study~\cite{2012LI} were insufficient to 
identify the ``droplet shape," as more than 500k water molecules were required for
that purpose~\cite{2013SAN}.
Within the framework of the direct approach, {\it ab initio} evaluations of the contact angle 
are still a challenge for AIMD, whereas classical MD involves a number of
case-by-case parameter fittings.
From the viewpoint of complexity of simulation procedures, 
our DFT-based indirect approach can be regarded as being simpler than the MD-based direct approach.

\subsection{Limitations in comparison with reality}
The present model achieves 
the maximum possible toward {\it the reality}
within the available computational feasibility, 
but there are several unsatisfactory 
factors to describe the reality as we can 
point out below. 
To obtain a quantitatively reliable 
estimate for the interaction energy 
between water and the surface, 
one would wonder that 
the larger coverages than one bilayer 
as well as structural disorder in the water layer
should be considered. 
To capture such factors, 
the larger simulation cells are required. 
The feasibility of practical {\it ab initio} 
calculations is, however, subject to limitations 
on simulation sizes. 
The computational cost typically scales as
the cube of the system size.~\cite{2016DUB} 
The available memory capacity is far 
smaller than that required to store the dynamic 
updates of such a larger system. 

\par
Not only in the present model, but also 
in all the current {\it ab initio} treatments, 
$\gamma_{LG}$ is approximately modelled by 
the surface energy of $Ih$-ice.~\cite{2009LAN,2016CAR}
Though there are several possible criticisms
against this approximations, 
this modelling is actually the {\it keystone} 
for the treatment, otherwise it becomes 
infeasible to describe the contact angle 
in the {\it ab initio} framework within the 
practical computational cost. 
At least within the current feasibility, 
we can either (a) compromise between 
the limitation of the modeling and 
the universality of {\it ab initio} framework 
to capture the trends with the same formalism, 
or (b) abandon the universality, just focusing on a 
specific system to pursue the descriptiveness of reality 
by using qualitative model-theoretical treatments. 

\par
As we mentioned in section ``Computational Methods'',
dispersion or vdW interactions 
is another point for further investigations. 
The exchange-correlation functionals capable of capturing
the dispersion interactions have gradually be underway, 
but they are typically a number of times heavier than 
the conventional ones in their computational cost 
as a result of the extra processing 
({\it e.g.}, non-locality) required.~\cite{2012JIR}
Hence, from the viewpoint of the computational cost, 
it is a trade-off between the direction toward 
capturing the disorder effects and that 
toward describing the dispersion interactions in detail. 
In addition, the current vdW-DF approach is
not appropriate for delocalized systems,~\cite{2017JYL_TJZa,2013TB_JGAa}
implying other approaches beyond DFT are
necessary for the present case.

\section{Conclusion}
In this study, we established two schemes
for estimating the water contact angle on the Cu (111) surface;
one is a regression-based interpolation of
the coverage-dependent contact angles
at the reference value of the coverage,
the other is the Boltzmann-averaged contact angle
over all the isolated clusters with different values of the coverage.
The Boltzmann-averaged estimate was found to
agree with the regression-based one within their error bars.
We found that experimental contact angles in literature remain ambiguous,
and the uncertainty of experimental contact angles
is larger than our two models' error bars.
Comparing our two models,
the regression-based estimate involves
the reference molecular coverage {\it a priori},
whereas the Boltzmann-averaged estimate
is more straightforward than the regression-based one.
This indicates the Boltzmann-averaged estimate
would be applicable to other metallic surfaces
without knowing a molecular coverage value {\it a priori}.

\section*{Acknowledgments}
The computations in this work were performed 
using the facilities of 
Research Center for Advanced Computing 
Infrastructure at JAIST. 
K.H. is grateful for financial support from 
the HPCI System Research Project (Project ID: hp190169 and hp200040) and 
MEXT-KAKENHI (JP16H06439, JP17K17762, JP19K05029, and JP19H05169)
and the Air Force Office of Scientific Research
(Award Number: FA2386-20-1-4036).
R.M. is grateful for financial support from 
MEXT-KAKENHI (JP19H04692 and JP16KK0097), 
FLAGSHIP2020 (project nos. hp190169 and hp190167 at K-computer), 
Toyota Motor Corporation, 
the Air Force Office of Scientific Research 
(AFOSR-AOARD/FA2386-17-1-4049;FA2386-19-1-4015), 
and JSPS Bilateral Joint Research Projects (with India DST). 
%
\bibliography{references}
\end{document}